\documentclass[twocolumn]{aastex631}

\usepackage{amsmath}
\usepackage{bm}

\usepackage{soul}

\shorttitle{SMBH Growth in Ellipticals}
\shortauthors{Farrah et al.}

\newcommand{\uhm}{Department of Physics and Astronomy, University of Hawai`i at M\=anoa, 2505 Correa Rd., Honolulu, HI, 96822, USA}

\begin{document}

\title{The assembly of supermassive black holes at $z<1$ in early-type galaxies from scaling relations}

\correspondingauthor{Duncan~Farrah}
\email{dfarrah@hawaii.edu}

\author[0000-0003-1748-2010]{D. Farrah}
\affiliation{\uhm}
\affiliation{Institute for Astronomy, University of Hawai‘i, 2680 Woodlawn Dr., Honolulu, HI 96822, USA}

\author[0000-0001-6970-7782]{A. Engholm}
\affiliation{\uhm}
\affiliation{Institute for Astronomy, University of Hawai‘i, 2680 Woodlawn Dr., Honolulu, HI 96822, USA}

\author[0000-0003-0917-9636]{E. Hatziminaoglou}
\affiliation{European Southern Observatory, Karl-Schwarzschild-Str. 2, 85748 Garching bei München, Germany}
\affiliation{Instituto de Astrofisica de Canarias (IAC), E-38205 La Laguna, Tenerife, Spain}
\affiliation{Universidad de La Laguna, Opto. Astrofisica, E-38206 La Laguna, Tenerife, Spain}

\author[0000-0003-0624-3276]{S. Petty}
\affiliation{NorthWest Research Associates, 3380 Mitchell Ln., Boulder, CO 80301, USA}

\author[0000-0001-8973-5051]{F. Shankar}
\affiliation{Department of Physics and Astronomy, University of Southampton, Highfield SO17 1BJ, UK}

\author[0000-0002-2612-4840]{A. Efstathiou}
\affiliation{School of Sciences, European University Cyprus, Diogenes Street, Engomi, 1516 Nicosia, Cyprus}

\author[0009-0002-6970-5247]{K. Ejercito}
\affiliation{\uhm}
\affiliation{Institute for Astronomy, University of Hawai‘i, 2680 Woodlawn Dr., Honolulu, HI 96822, USA}

\author[0000-0001-7685-158X]{K. Jones}
\affiliation{\uhm}

\author[0000-0002-3032-1783]{M. Lacy}
\affiliation{National Radio Astronomy Observatory, Charlottesville, VA, USA}

\author{C. Lonsdale}
\affiliation{National Radio Astronomy Observatory, Charlottesville, VA, USA}

\author[0000-0001-6139-649X]{C. Pearson}
\affiliation{RAL Space, STFC Rutherford Appleton Laboratory, Didcot, Oxfordshire OX11 0QX, UK}
\affiliation{The Open University, Milton Keynes MK7 6AA, UK}
\affiliation{Oxford Astrophysics, University of Oxford, Keble Rd, Oxford OX1 3RH, UK}

\author[0000-0003-1704-0781]{G. Tarl\'e}
\affiliation{Department of Physics, University of Michigan, 450 Church St., Ann Arbor, MI, 48109, USA}

\author[0000-0001-8156-6281]{R. A. Windhorst} 
\affiliation{School of Earth and Space Exploration, Arizona State University, Tempe, AZ 85287-6004, USA}

\author[0000-0002-9149-2973]{J. Afonso}
\affiliation{Instituto de Astrof\'{i}sica e Ci\^{e}ncias do Espa\c co, Universidade de Lisboa, Portugal}
\affiliation{Departamento de F\'{i}sica, Faculdade de Ci\^{e}ncias, Universidade de Lisboa, Portugal}

\author[0000-0002-9548-5033]{D. L. Clements}
\affiliation{Imperial College London, Blackett Laboratory, Prince Consort Road, London, SW7 2AZ, UK}

\author[0000-0002-6917-0214]{K. S.~Croker}
\affiliation{School of Earth and Space Exploration, Arizona State University, Tempe, AZ 85287-6004, USA}
\affiliation{\uhm}

\author[0000-0002-5206-5880]{L. K. Pitchford}
\affiliation{Department of Physics and Astronomy, Texas A\&M University, College Station, TX, USA} 
\affiliation{George \& Cynthia Woods Mitchell Institute for Fundamental Physics and Astronomy, Texas A\&M University, College Station, TX, USA}

\begin{abstract}
The assembly of supermassive black hole (SMBH) mass  ($M_{\bullet}$) and  stellar mass ($M_{*}$) in galaxies can be studied via the redshift evolution of the $M_{\bullet}-M_{*}$ relation, but the ways in which selection bias and physical assembly channels affect this evolution are uncertain. To address this, we compare the $M_{\bullet}-M_{*}$ relation for local massive ($M_{*}>10^{10.5}$M$_{\odot}$) quiescent early-type galaxies (ETGs) to that for massive ETGs hosting active galactic nuclei (AGN) at $z\sim0.8$. The restrictions on stellar mass and galaxy type limit the assembly channels that may connect the two relations. For the local sample we find $\log(M_{\bullet}) = 8.80 + 1.10(\log{M_{*}-11})$, in line with prior work.  For the $z\sim0.8$ sample we find a bias-corrected relation: $\log(M_{\bullet}) = 7.80 + 1.25(\log{M_{*}-11})$.  We show, however, that this relation depends on the stellar and SMBH mass functions used to compute the selection bias, the virial relation, the virial factor, and the active fraction, which together introduce uncertainty of up to $\sim0.6$\,dex in the $z\sim0.8$ relation. Adopting reasonable choices of these parameters then our $z\sim0.8$ relation lies above that for $z\sim0$ AGN by $\sim0.5$\,dex, but below our $z\sim0$ ETG relation by $0.4-1$\,dex in SMBH mass. We discuss possible sources of this offset, including further bias corrections, `downsizing" in SMBH mass assembly, and preferential SMBH growth. Our results highlight the need to reduce uncertainties from selection and measurement bias in SMBH and stellar masses at all redshifts. 
\end{abstract}


\section{Introduction}
Among galaxies at $z\sim0$, scaling relations exist between SMBH mass and some properties of their host galaxies \citep{gebh00,feme04,mccma13,kormendyho13,zhao21}. Potential origins of these scaling relations include feedback between the SMBH and star formation in its host galaxy \citep{fabian12,silk13,weinberger17}, a common fuel reservoir for SMBH accretion and star formation \citep[e.g.][]{ni21}, and galaxy-galaxy mergers \citep[e.g.][]{hirsch10}.  

Studies of the evolution of galaxy scaling relations with redshift can, in principle, diagnose the processes that generate them.  To date however, studies of this redshift evolution have produced conflicting results. Some find relatively more massive SMBHs at higher redshift \citep{decarli10,merloni10,ding20,zhang23}.  Others find more massive SMBHs at lower redshifts \citep{ueda18}, or no clear evidence for evolution \citep{schramm13,suh20,lijen21,tanaka24,cloonan24}.   The origin of these discrepancies may lie in one or more of selection bias, measurement bias, and physical mass change channels. 

Progress on determiming which factors influence the redshift evolution of galaxy scaling relations can be made by restricting  studies to specific galaxy types, since this limits the processes that must be considered. ETGs at $z<1$ with total stellar masses $M_{*}\gtrsim10^{10.5}M_{\odot}$ are suited to this purpose as they emerge en masse onto the red-sequence at this epoch,  on which they evolve fairly passively, and because the $M_{\bullet}-M_{*}$ relation likely does not strongly deviate from a log-linear form at $M_{*}\gtrsim10^{10.5}M_{\odot}$ (\citealt{sijacki15}, though see \citealt{habouzit21}).  A recent study that took this approach \citep[][F23 hereafter]{farrah23ellip}  compared the  $M_{\bullet}-M_{*}$ relation in local quiescent ETGs to that for AGN hosted in ETGs at $z\gtrsim0.8$. They found evidence for relatively less massive SMBHs in ETGs at  $z\sim0.8$ compared to locally,  by a factor of about seven. However, while F23 made bias corrections,  their selection bias correction was not based on a direct estimate from galaxy populations. They also used a novel method to compare the high- and low-redshift samples, via shifting populations in the $(M_{\bullet},M_{*})$ plane, rather than the usual approach of deriving analytic scaling relations, which limited comparisons to prior works. 

In this paper, we improve on the analysis in F23. We focus on two samples; a low-redshift ($z<0.2$) quiescent ETG sample and a high-redshift ($z\sim0.8$) sample of AGN in ETG hosts, selected to represent ETGs about to emerge onto the red-sequence. We compare the $M_{\bullet}-M_{*}$ relations of the two samples to each other to examine the assembly of ETGs since $z\sim1$. In doing so, we explore the impact of different selection and measurement bias corrections.  \S\ref{sec:sm_sel} presents the two samples. \S\ref{sec:allbiases} describes the selection and measurement bias corrections. \S\ref{sec:res} presents our results, including the low-redshift  $M_{\bullet}-M_{*}$ relation, the high-redshift observed and intrinsic $M_{\bullet}-M_{*}$ relations, and the processes that may align them.  We discuss these results in \S\ref{sec:discuss} and present our conclusions in \S\ref{sec:conc}. We assume \mbox{$H_0 = 70$\,km\,s$^{-1}$\,Mpc$^{-1}$}, \mbox{$\Omega = 1$}, and \mbox{$\Omega_{\Lambda} = 0.7$}. We convert literature data to this cosmology and a \citet{kroupa03} initial mass function (IMF) where necessary. For ease of comparison with past work we use $\mu = \log_{10}M_{\bullet}$ and  $s = \log_{10}M_{*}$. All uncertainties on derived parameters are $2\sigma$ (95\%).

\begin{deluxetable}{lrrr}
\tablehead{
\colhead{Name} & \colhead{$z$} & \colhead{$M_{\bullet}$} & \colhead{$M_{*}$} \\
\colhead{}     & \colhead{}    & \multicolumn{2}{c}{$\log$M$_{\odot}$}
}
\tablecaption{Additional low-redshift ETGs (\S\ref{sec:sm_sel}).\label{tbl:lowz_sample}}
\startdata
Abell 1201 &  0.169 & $10.52^{+0.09}_{-0.11}$ & $12.25\pm0.30$ \\
UGC 2698   &  0.021 & $9.39^{+0.12}_{-0.17}$  & $11.56\pm0.15$ \\
NGC 708    & 0.016 & $10.00^{+0.11}_{-0.14}$  & $11.45\pm0.30$ \\
NGC 1272    & 0.013 & $9.70^{+0.20}_{-0.40}$  & $11.90\pm0.30$ \\
NGC 2832   & 0.023 & $9.77^{+0.13}_{-0.18}$  & $12.08\pm0.30$ \\
NGC 3258   & 0.009 & $9.35^{+0.05}_{-0.05}$  & $11.72\pm0.30$ \\
NGC 4281   & 0.008 & $8.69^{+0.02}_{-0.02}$  & $10.88\pm0.50$ \\
\enddata
\tablerefs{\citealt{night23,cohn21,denicola24,saglia24,greene19,boizelle19,thater19}}
\end{deluxetable}

\begin{deluxetable*}{rccccclccccccl}
\tablehead{
\colhead{$\#$} &
\colhead{$\Phi^{*}$} &
\colhead{$s^{*}$} & 
\colhead{$\alpha_{*}$} &  
\colhead{$\Phi^{2,*}$} & 
\colhead{$\alpha_{2,*}$} & 
\colhead{Ref} &
\colhead{$\Phi^{\bullet}$} & 
\colhead{$\mu^{\bullet}$} & 
\colhead{$\alpha_{\bullet}$} &  
\colhead{$\beta_{\bullet}$} &  
\colhead{$\lambda^{\lambda}$} &  
\colhead{$\alpha_{\lambda}$} & 
\colhead{Ref} 
}
\tablecaption{The SMF (Schechter) and BHMF/ERDF (modified Schechter \& Schechter) combinations. Listed are the normalizations, break parameters (mass or Eddington ratio) and slope(s).  The M21 SMFs are their intrinsic $0.75<z<1.25$ versions. The W16 SMF is for $z\sim0$ with sSFRs of $10^{-11} - 10^{-12}$yr$^{*1}$.  The W23 SMFs are for the 1.3deg$^{2}$ COSMOS field at $0.8<z<1.1$. The BHMFs/ERDFs from S15 are at $z\sim0.8$. Set \#1 is our default (\S\ref{sec:hizsel}).\label{tbl:paramcombs}}
\startdata
1 & -3.01 & 10.86 & -1.37 & --    &  --       & M21 (single)  & -4.88 & 8.06 & -1.19  & 0.57 & -1.02 & -1.09 & S15 (SDSS) \\  
2 & -2.67 & 10.51 &  0.08 & -3.07 & -1.49 & M21 (double) & -4.88 & 8.06 & -1.19  & 0.57 & -1.02 & -1.09 & S15 (SDSS)  \\  
3 & -2.23 & 10.64 & -0.52 & --    &  --       & W16    & -4.88 & 8.06 & -1.19  & 0.57 & -1.02 & -1.09 & S15 (SDSS) \\  
4 & -5.96 & 10.89 & -2.01 & -3.02 & -0.47 & W23 (quiescent) & -4.88 & 8.06 & -1.19  & 0.57 & -1.02 & -1.09 & S15 (SDSS)  \\  
5 & -3.08 & 11.02 & -1.32 & -3.18 & -0.63 & W23 (total) & -4.88 & 8.06 & -1.19  & 0.57 & -1.02 & -1.09 & S15 (SDSS) \\  
6 & -3.08 & 11.02 & -1.32 & -3.18 & -0.63 & W23 (total) & -5.32 & 9.09 & -1.50  & 0.96 & -1.19 & -0.29 & S15 (Comb.)  \\  
\enddata
\end{deluxetable*}

\section{Sample Selection}\label{sec:sm_sel}
To assemble the low-redshift sample, we start with the low-redshift ETGs in F23. Their selection criteria are: an early-type morphology, a total stellar mass of M$_{*}>10^{10.5}$M$_{\odot}$, an SMBH mass measured via stellar dynamical modelling, and no evidence for an AGN, pseudobulge, or bar. We exclude NGC 5018 as its dusty center may make stellar mass estimates difficult. We add seven objects that were either presented recently, or have an SMBH mass measured via gas-dynamical modelling (Table \ref{tbl:lowz_sample}) since gas- and stellar-dynamical modelling give similar SMBH masses \citep[e.g.][]{waters24}.  The final sample comprises 57 objects. 

To assemble the high-redshift sample, we follow the procedure in F23. In F23, the WISE-selected AGN catalog of \citet{barrows21} was merged with the Sloan Digital Sky Survey (SDSS) DR14 quasar catalog of \citet{raks20} to form the parent catalog. AGN were then selected with $0.8<z<0.9$, a total stellar mass of M$_{*}>10^{10.5}$M$_{\odot}$, an ETG host spectral energy distribution (SED), AGN reddening of $E(B-V)<0.2$, and SFRs at least a factor of three below the SFR-M$_{*}$ main sequence. The redshift range places the sample near the end of the emergence of the ~red-sequence, and is low enough that the  \citet{barrows21} catalog is reasonably complete in stellar mass to M$_{*}=10^{10.5}$M$_{\odot}$. The narrow redshift range also means the sample can be treated as luminosity-limited, which is necessary for the selection bias correction in \S\ref{sec:hizsel}. We make two changes to the procedure in F23. First, we update to the \citet{wushen22} quasar catalog, which is based on SDSS DR16. The resulting sample is almost identical to that obtained using the \citet{raks20} catalog. Second, since the bias model in \S\ref{sec:hizsel} assumes a luminosity-limited sample, we find the luminosity limit of the sample, $10^{45.5}$ ergs s$^{-1}$, using the bolometric luminosities from   \citet{barrows21} of all quasars at $0.8<z<0.9$ in \citet{wushen22}. We remove the quasars below this limit, to leave a final sample of 287 objects. We do not consider the other high-redshift samples in F23, or other literature samples, as they lack the combination of homogeneous selection, a narrow redshift range, large sample size, and homogneously measured AGN and host properties.

\section{Bias corrections}\label{sec:allbiases}
We aim to compare the intrinsic $M_{\bullet}$ to intrinsic $M_{*}$ relation derived from the low-redshift sample to the same relation derived from the high-redshift sample immediately after their AGN phase completes and they settle into quiescence.  This requires several bias corrections, which we detail below.

\begin{figure*}
\begin{center}
\includegraphics[width=0.98\linewidth]{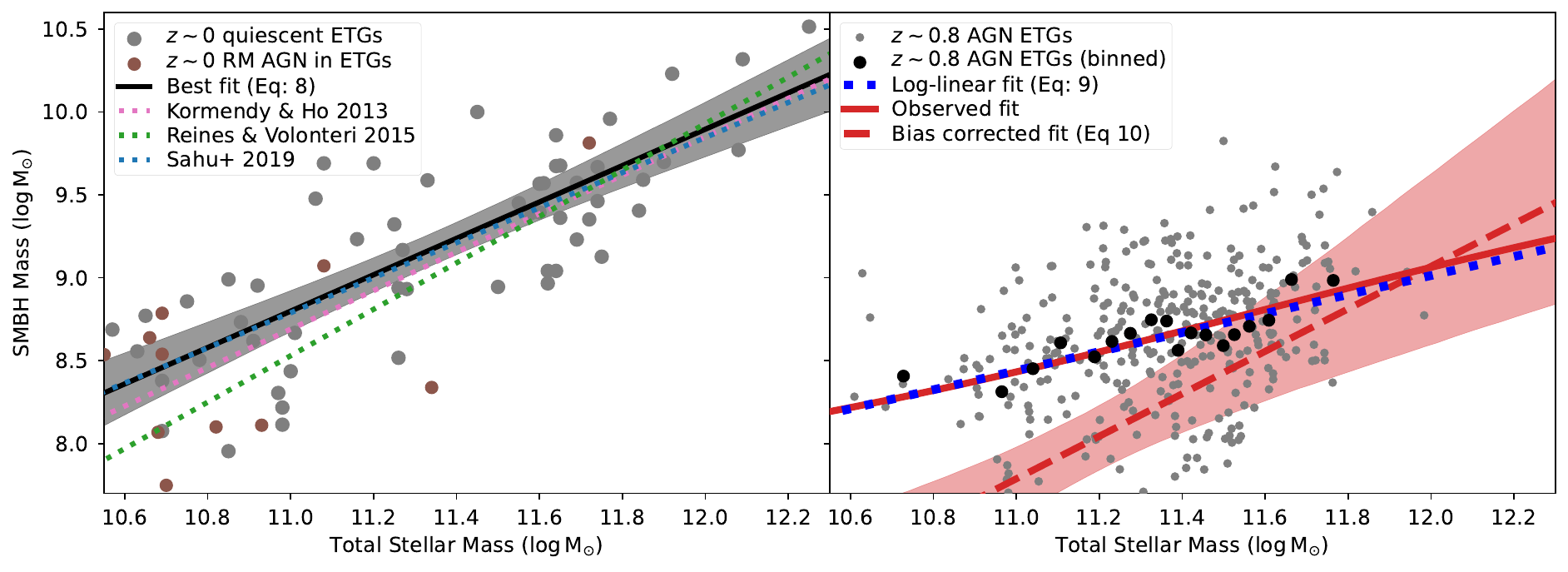} 
\caption{{\itshape Left:} The low-redshift sample in the $M_{\bullet}-M_{*}$ plane with our best fit (Equation \ref{eq:simplelocal}). Also plotted are the relations for ETGs from \citet{kormendyho13,rein15,sahu19}, and the  compilation of AGN in ETG hosts at $z\sim0$ from F23. {\itshape Right:} The high-redshift sample in the $M_{\bullet}-M_{*}$ plane with our log-linear fit (Equation \ref{eq:simplehighz}) and selection-bias corrected fit (Equation \ref{eq:selectionhiz}) and observed fits from the model in \S\ref{sec:hizsel}. The shaded regions are the $2\sigma$ confidence intervals.}
\label{fig:lhplanes}
\end{center}
\end{figure*}

\subsection{Preliminary considerations}\label{sec:prelimoverall}

\subsubsection{Accretion bias}\label{sec:prelim1}
Since the high-redshift sample are within an AGN phase, SMBH growth via accretion before they transition to quiescence is inevitable, and must be accounted for.  F23 do so by increasing the individual SMBH masses by an amount drawn from: $\Delta\log M_{\bullet}=U(0, 0.1)$ (their appendix B2). Here, our approach makes this accretion degenerate with other SMBH growth channels. So, we conservatively assume a fixed correction equal to the maximum in F23, of 0.1 dex.

 \subsubsection{Stellar mass biases}\label{sec:prelim2}
 For the stellar masses of the low- and high-redshift samples, two corrections must be considered.  First is measurement bias. Second is Eddington bias, in which systems with smaller stellar masses are more likely to scatter upwards due to measurement error, than are systems with larger stellar masses to scatter downwards.   
 
For the low-redshift sample, we estimate the stellar measurement bias by comparing to the objects in common with  \citet{watkins22}, who estimate stellar masses via integrated $3.6\mu$m and $4.5\mu$m photometry. Their stellar masses are $\sim0.1$\,dex lower than ours.  Since it is unclear which of the measurement methods is closer to the truth, we do not apply a measurement bias correction.  The level of Eddington bias in the low-redshift sample is also likely small, so for simplicity we do not correct for this either. We note, though, that applying either correction would increase the offset between the low- and high-redshift $M_{\bullet}-M_{*}$ relations in \S\ref{sec:res}.

For the high-redshift sample, measurement bias may arise because the stellar masses of the high- and low-redshift samples are measured in different ways; SED decomposition and near-infrared spatial profile fitting, respectively. This bias is, however, challenging to estimate. F23 find that measurement bias lowers the stellar masses of the high-redshift sample by $0.2-0.3$\,dex, though this could also arise from aligning two-dimensional inhomogeneous samples. \citet{li21rmsloan} find that stellar masses from SED decomposition are $\sim0.1$\,dex lower than those  from profile fitting, though their sample is not directly comparable to ours. A measurement bias of $\sim0.1-0.2$\,dex low thus seems plausible, though a bias outside this range in either direction cannot be ruled out. The Eddington bias in the high-redshift sample is also difficult to estimate, as it will depend on the parameters in our selection bias model (\S\ref{sec:redvary}). Self-consistently investigating this bias is beyond the scope of this paper.  However, we argue that the Eddington bias is small. Using the combinations of mass function parameters in Table \ref{tbl:paramcombs} then this bias is negligible at $M_{*}\lesssim10^{10.7}M_{\odot}$, rising to $0.1-0.2$\,dex at $M_{*}\sim10^{11.5}M_{\odot}$. The measurement bias and the Eddington bias for the high-redshift samples are thus of comparable magnitude but opposite sign. We therefore make the simplifying assumption that these corrections cancel each other, and so do not apply either.  We note that the magnitude of either is smaller than the uncertainty in the selection bias correction (\S\ref{sec:redvary}).

\subsubsection{SMBH measurement bias}\label{sec:methvir}
The SMBH masses of the two samples are measured using different methods - dynamical modelling for the low-redshift sample, and \ion{Mg}{2}$\lambda$2799\AA\ single-epoch virial (SEV) masses for the high-redshift sample. Following \citet{thater22}, we assume negligible measurement bias in the low-redshift SMBH masses (though see \citealt{thater23}). For the high-redshift sample, there exist several SEV calibrations, which can differ by $\sim0.3$\,dex.  We adopt the \citet{shen11cat} calibration (S11 hereafter), as it is used to compute the SMBH mass functions in \S\ref{sec:hizsel}. The S11 relation has a scatter of $\sigma_{V}\simeq0.4$\,dex. We explore other SEV choices in  \S\ref{sec:redvary}.

\subsection{Selection bias}\label{sec:hizsel}
As they are AGN, the SMBH masses of the high redshift sample may  be biased high with respect to the SMBH population. In F23 this was accounted for via a correction drawn from $\Delta\log M_{\bullet}=U(-0.3 , -0.1)$. We improve on this treatment by adopting the bias model of \citet{schwi11}. The model assumes an intrinsic mean $M_{\bullet}-M_{*}$ relation: 

\begin{equation}\label{intmus}
\mu = a + bs
\end{equation}

\noindent with scatter $\sigma_{i}$. If $\mu$ and $s$ are log-normally distributed, then the probability of obtaining $\mu$, given $s$, is:

\begin{equation}\label{eq:gfunc}
g(\mu | s) = \frac{1}{\sqrt{2\pi}\sigma_{i}}exp\left(\frac{(\mu - a - bs)^2}{2\sigma_{i}^2}\right)
\end{equation}

\noindent Given an intrinsic stellar mass function (SMF) $\Phi_{*}(s)$, the {\itshape total} SMBH mass function (BHMF) $\Phi_{\bullet,tot}(\mu)$, is:

\begin{equation}\label{eq:totalbhmf}
\Phi_{\bullet,tot}(\mu) = \int g(\mu | s) \Phi_{*}(s) ds
\end{equation}

\noindent The selection bias is quantified by comparing Equation \ref{eq:totalbhmf} to the intrinsic active BHMF, $\Phi_{\bullet,act}(\mu)$, itself moderated by selection biases in the sample relative to all active SMBHs. For a sample luminosity limited at $L_{lim}$ the SMBH selection function, $\Omega (\mu)$, is: 

\begin{equation}
\Omega(\mu) = p_{act}\int_{L_{lim}}^{\infty}p(L,\mu)dL
\end{equation}

\noindent in which $p_{act}$ is the fraction of SMBHs that are active, and $p(L,\mu)$ is the probability of finding an SMBH at a point in the ($L, \mu$) plane. The active fraction is:

\begin{equation}\label{eq:actfrac}
p_{act} = \frac{\Phi_{\bullet,act}(\mu)}{\Phi_{\bullet,tot}(\mu)}
\end{equation}

\noindent and $p(L,\mu)$ is:

\begin{equation}
p(L,\mu) = \frac{\Phi_{\lambda}(\lambda)}{\int \Phi_{\lambda}(\lambda) d\lambda}
\end{equation}

\noindent in which $\Phi_{\lambda}(\lambda)$ is the Eddington ratio distribution function (ERDF). The relation between stellar mass and observed SMBH mass is then:

\begin{equation}\label{obsmus}
\langle \mu \rangle (s) = \frac{\int \mu\Omega(\mu)g(\mu | s) d\mu}{\int \Omega(\mu)g(\mu | s) d\mu}
\end{equation}

\noindent in which $\langle \mu \rangle (s)$ is the mean of the logarithm of the SMBH masses, rather than the logarithm of their mean. The difference between Equations \ref{obsmus} and \ref{intmus} is the selection bias in the sample. 

Equation \ref{obsmus} requires the intrinsic SMF and the active BHMF \& ERDF. We adopt functions from studies based on samples close to ours. For the SMF we use the single-Schechter intrinsic SMF at $0.75<z<1.25$ from \citet[][M21 hereafter]{mcleod21}. This SMF is based on a near-infrared selected sample from multiple fields spanning $\sim3$ deg$^{2}$, and has good sampling of stellar masses at $>10^{11}$M$_{\odot}$. Their use of a Chabrier IMF should not introduce significant bias as it is close to a Kroupa IMF. For the active BHMF \& ERDF we adopt modified Schechter and Schechter forms, respectively, from \citet[][S15 hereafter]{schulze15}. Their ERDF assumes an $M_{\bullet}$-dependent break Eddington ratio (their equations 15 and 16). We select their ``SDSS" parameters at $z\sim0.8$, since our sample is drawn from the SDSS. This parameter set is listed in Table \ref{tbl:paramcombs}. 

The SMBH masses of the low-redshift sample may also be biased high, as the method used to measure them requires that the SMBH sphere of influence be resolved. For $M_{*}>10^{10.5}$M$_{\odot}$ it has been argued that this bias is up to $\sim0.4$\,dex \citep{shank16a,shank20,carrero20}. We consider this possibility in \S\ref{sec:res}.

\begin{figure*}
\begin{center}
\includegraphics[width=0.98\linewidth]{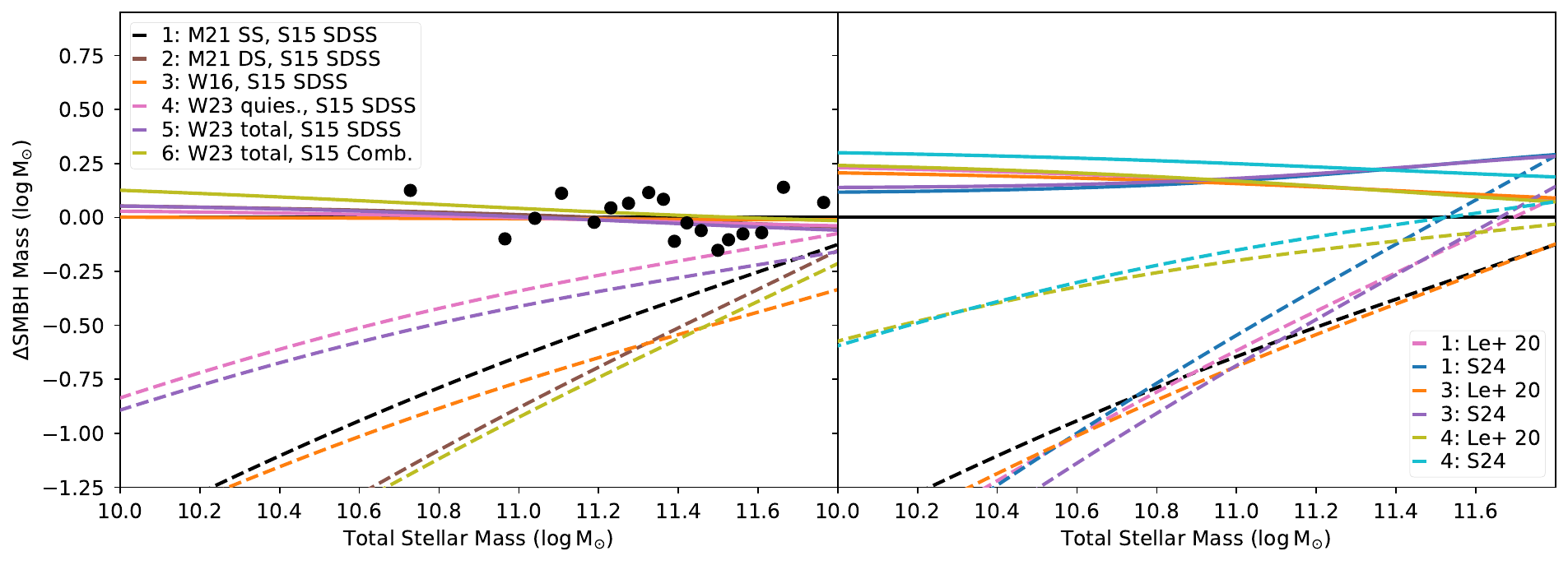} 
\caption{The impact on the high-redshift observed (solid) and intrinsic (dashed) fit of ({\itshape left}) the SMF, BHMF, and ERDF parameters (Table \ref{tbl:paramcombs}), and ({\itshape right}) the SEV calibrator (\S\ref{sec:redvary}). The \citet{zuo2015} results are close to the \citet{le20mg2} results and are omitted for clarity. In both panels, the fits are shown relative to the observed fit using parameter set \#1. The data are not plotted in the right panel since they depend on the SEV calibrator.}
\label{fig:vsbtest1}
\end{center}
\end{figure*}

\section{Results}\label{sec:res}

\subsection{The low- and high-redshift $M_{\bullet}-M_{*}$ relations}\label{sec:resint}
We fit a function of the form in Equation \ref{intmus} to the low-redshift sample\footnote{All model fitting was performed within Python 3.10.1, using Dynesty v2.1.2 \citep{dynesty2020}, with a random seed of 1001.}. This gave a mean relation:

\begin{equation}\label{eq:simplelocal}
\mu = 8.80^{+0.12}_{-0.12} + 1.10^{+0.21}_{-0.21} (s - 11)
\end{equation}

\noindent (Figure \ref{fig:lhplanes}, left), with $\sigma_{obs} = 0.37^{+0.08}_{-0.06}$ (this includes intrinsic scatter, observational uncertainty, and scatter from calibration uncertainties in the stellar and SMBH masses). These results are in agreement with prior work \citep{kormendyho13,rein15,sahu19}. As verification, we compare in Figure \ref{fig:lhplanes}, left to the compilation of AGN in ETG hosts at $z\sim0$ from F23 with SMBH masses from reverberation mapping or the Event Horizon Telescope. This compilation is also consistent with our fit.

For the high-redshift sample; we first give a direct comparison to the low-redshift sample by fitting a function of the form in Equation \ref{intmus}. This yields:

\begin{equation}\label{eq:simplehighz}
\mu = 8.46^{+0.08}_{-0.08} + 0.53^{+0.18}_{-0.17} (s - 11)
\end{equation}

\noindent  (Figure \ref{fig:lhplanes}, right), with $\sigma_{obs} = 0.39^{+0.04}_{-0.03}$. This scatter is comparable to that in the S11 relation. It is possible that the selection of our host galaxies as ETGs reduces this scatter, but we do not explore this further. 

Next, we derive the intrinsic mean high-redshift $M_{\bullet} - M_{*}$ relation by applying the selection bias model in \S\ref{sec:hizsel} to the high-redshift sample.  To reduce computation time, we bin the sample in 18 bins of stellar mass, equally divided in object number. We then evaluate the log-likelihood by comparing the predictions from Equation \ref{obsmus} to these binned data. We adopt parameter set \#1 from Table \ref{tbl:paramcombs} and the S11 SEV calibration.  Since $p_{act}$ can lie above unity for arbitrary choices of parameters in Equation \ref{eq:actfrac}, and because studies that constrain $p_{act}$ are to some extent model-dependent \citep{aversa15,delve20} we impose a loose constraint: $10^{-4} < p_{act} < 0.2$ at any $M_{\bullet}$. Since the scatter in Equation \ref{intmus}, $\sigma_{i}$, may be affected by the scatter in the SEV calibration, $\sigma_{V}$, we model the scatter in Equation \ref{intmus} as $\sigma_{i} = \sqrt{\sigma_{t}^{2} + \sigma_{V}^{2}}$, where $\sigma_{t}$ is the true intrinsic scatter\footnote{The scatter that goes into the log-likelihood, $\sigma_{ll}$, is not the same as $\sigma_{i}$ or $\sigma_{t}$ - $\sigma_{ll}$ does not enter Equation \ref{eq:gfunc}. or Equation \ref{obsmus}. Instead, $\sigma_{ll}$  is the scatter among the {\itshape binned} points in Figure \ref{fig:lhplanes}, right and only appears in the log-likelihood.  It is found to be small: $\sigma_{ll}=0.11^{+0.02}_{-0.04}$.}. The fit yields:

\begin{equation}\label{eq:selectionhiz}
\mu = 7.80^{+0.18}_{-0.24} + 1.25^{+0.68}_{-0.47} (s - 11)
\end{equation}

\noindent (Figure \ref{fig:lhplanes}, right), with $\sigma_{t} = 0.24^{+0.18}_{-0.38}$. The slope is consistent within $1\sigma$ with the slope in Equation \ref{eq:simplelocal}.  There is degeneracy between $\sigma_{t}$ and both $a$ and $b$; a Kendall-tau test yields $(\tau,p) = (-0.27,<0.01)$ and $(\tau,p) = (0.44,<0.01)$, respectively. The effects are, however, small -  increasing $\sigma_{t}$ by $0.2$ decreases $a$ by 0.08 and increases $b$ by 0.25. Comparing Equation \ref{eq:selectionhiz} to the observed mean relation (Equation \ref{obsmus}) implies stellar mass dependent selection bias in the sample. The level of bias is in agreement with other high-redshift luminous AGN \citep{shen08,schwi11}.

\subsection{Effect of parameter choices}\label{sec:redvary}
We first consider the impact on the high-redshift selection-bias fit of the SMF, and active BHMF \& ERDF.  Our choice of  parameter set \#1 is based on the selection of the sample at $z\sim0.8$ from SDSS fields. Since, however, these parameters are not measured for our sample in particular, other choices are possible.  To explore the sensitivity of our results to this choice, we consider the following SMFs: the double-Schechter intrinsic SMF at $z\sim0.8$ from M21, the quiescent and total SMFs at $z\sim0.8$ from \citet[][W23 hereafter]{weaver23smf}, and the $z\sim0$ SMF for quiescent SDSS galaxies from \citet[][W16 hereafter]{weigel16}. The W23 and W16 SMFs may not be as well-matched to our sample as those from M21 as they are either constructed in smaller fields, or at lower redshift, but they shoud serve as comparisons. For the active BHMF and ERDF there are few studies in our redshift range, so we draw on S15 again, who present a ``Combined" BHMF/ERDF derived using several datasets, including the SDSS.   

We fit our high-redshift sample with combinations of these SMFs, BHMFs, and ERDFs, but found limited preference for one over another. Replacing our default SMF with any of the alternative SMFs gave comparable quality fits.  For the active BHMF/ERDF, the ``SDSS" version gave better quiality fits in most (but not all) cases.  A selection of parameters that give acceptable fits are listed in Table \ref{tbl:paramcombs} and plotted in Figure \ref{fig:vsbtest1}, left. While these combinations give comparable observed relations, they can give markedly different intrinsic relations.  For example, using parameter set \#4 gives:

\begin{equation}\label{eq:selectionhizp4}
\mu = 8.09^{+0.21}_{-0.20} + 0.95^{+0.42}_{-0.34} (s - 11)
\end{equation}

\noindent with $\sigma_{t} = 0.09^{+0.25}_{-0.40}$.

We explored if our high-redshift sample can, in concert with the selection bias model, constrain the SMF, BHMF, and ERDF. To do so, we allowed their parameters to vary with wide uniform priors. This yielded no useful constraints. We found $6.7 \lesssim a \lesssim 8.8$, $0.5 \lesssim b \lesssim 2.2$,  and $\sigma_{t}<1$. The single-Schechter SMF, BHMF and ERDF parameters were constrainded to lie within an order of magnitude of the parameter set \#1 values. 

Next, we examined the impact of adopting a different SEV calibration to that of S11. We considered three alternatives. \citet{zuo2015} present a calibration for quasars at $z\sim3.5$. \citet{le20mg2} consider a low-redshift sample over $\sim5$\,dex in luminosity. \citet[][S24 hereafter]{shen23rmkey} present a relation based on the SDSS reverberation mapping program. A caveat to this analysis is that our adopted BHMFs/ERDFs are computed using the S11 calibrator. Changing the SEV calibrator will change the parameters of the BHMF/ERDF.  Self-consistently recomputing the BHMF/ERDF for the alternative SEV calibrators is beyond the scope of this paper, so we take a simpler approach that should still be indicative. We increase (decrease) $\mu^{*}$ ($\lambda^(\lambda)$) by the average increase in the SMBH masses using the alternate SEV calibrators.  For \citet{zuo2015} and  \citet{le20mg2}  this is 0.15 dex, and for S24 this is 0.25 dex. 

The results are shown in Figure \ref{fig:vsbtest1}, right. Different SEV calibrator can affect the derived intrinsic relation, even for the same SMF, BHMF, and ERDF. For example, using parameter set \#1 with the S24 calibrator gives an intrinsic relation:

\begin{equation}\label{eq:selectionhizset1S24}
\mu = 7.90^{+0.19}_{-0.22} + 1.67^{+0.49}_{-0.51} (s - 11)
\end{equation}

\noindent with $\sigma_{t}=0.24^{+0.23}_{-0.38}$ (cf Equation \ref{eq:selectionhiz}). With parameter set \#4 the S24 calibrator gives:

\begin{equation}\label{eq:seelctionhizs24set4}
\mu = 8.29^{+0.17}_{-0.26} + 0.96^{+0.30}_{-0.40} (s - 11)
\end{equation}

\noindent with $\sigma_{t}=0.08^{+0.20}_{-0.25}$ (cf Equation \ref{eq:selectionhizp4}). 

Finally, motivated by the possibility that the intrinsic  low-redshift $M_{\bullet} - M_{*}$ relation is of log-polynomial form \citep{shank16a}, we test if the high-redshift intrinsic relation is of log-polynomial form. To do so, we modify Equations \ref{intmus} and \ref{eq:gfunc} to read:

\begin{equation}\label{eq:intpoly}
\mu = a + bs - cs^2 - ds^3
\end{equation}

\begin{equation}
g(\mu | s) = \frac{1}{\sqrt{2\pi}\sigma_{i}}exp\left(\frac{(\mu - a - bs+cs^2 + ds^3)^2}{2\sigma_{i}^2}\right)
\end{equation}

\noindent respectively, and then fit for the coefficients in Equation \ref{eq:intpoly}.  We found no significant change in fit quality. Coefficients $a$ and $b$ were consistent with their original values, though with larger errors. Coefficient $c$ was consistent with zero, with large uncertainties. There was a hint that $d<0$ but only at $2\sigma$ significance. We thus find no evidence preferring a log-polynomial relation, though we cannot rule it out.

\begin{figure*}
\begin{center}
\includegraphics[width=0.98\linewidth]{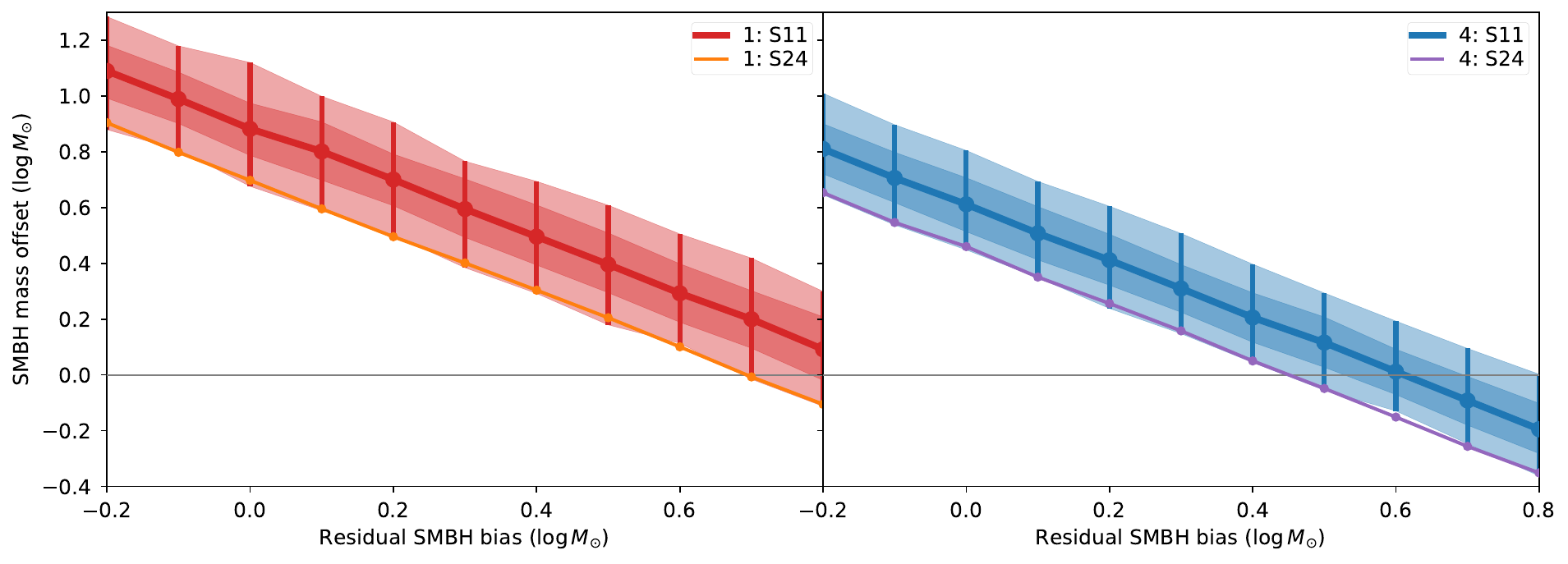}  
\caption{The SMBH mass offset needed to align the low- and intrinsic high-redshift relations for  parameter sets \#1 \& \#4, and the S11 and S24 SEV calibrators, as a function of residual SMBH mass bias (\S\ref{sec:resglo}).  The shaded regions show the 1$\sigma$ and 2$\sigma$ confidence intervals.}
\label{fig:offsetresidual}
\end{center}
\end{figure*}

\subsection{Redshift evolution of the $M_{\bullet}-M_{*}$ relation}\label{sec:resglo}
We here determine the offset in SMBH mass needed to align our high-redshift intrinsic $M_{\bullet}-M_{*}$ relation with our low-redshift mean relation.  To do so, we modify our analysis to draw the intrinsic slope of the high-redshift relation from the slope posterior of the low-redshift relation, rather than treating it as a free parameter. We then fit for the SMBH mass offset needed to align the relations using the selection bias model in  \S\ref{sec:hizsel} and subsequent application of the accretion bias in \S\ref{sec:prelim1}.  Since there may be `residual' SMBH bias arising from effects we have not corrected for, we perform several such fits, each for an amount of fixed residual SMBH bias between -0.2 and 0.8 dex. Following the results in \S\ref{sec:redvary}, we restrict this analysis to parameter sets \#1 and \#4, and the S11 and S24 SEV calibrators. 

The results are shown in Figure \ref{fig:offsetresidual}. With parameter set \#1, the S11 (S24) SEV calibrator, and  no residual SMBH mass bias the SMBH offset between the high and low redshift relations is $\Delta M_{\bullet}=0.88^{+0.24}_{-0.21}$\,dex, $\sigma = 0.17^{+0.21}_{-0.24}$  ($\Delta M_{\bullet}=0.69^{+0.20}_{-0.16}$\,dex, $\sigma = 0.11^{+0.20}_{-0.20}$).  With parameter set \#4 and the S11 (S24) SEV calibration we recover $\Delta M_{\bullet}=0.61^{+0.20}_{-0.16}$\,dex, $\sigma = 0.09^{+0.17}_{-0.13}$ ($\Delta M_{\bullet}=0.46^{+0.28}_{-0.25}$\,dex, $\sigma = 0.09^{+0.20}_{-0.13}$).  There is degeneracy between $\Delta M_{\bullet}$ and $\sigma$ -  a Kendall-tau test yields $(\tau,p) = (0.12,<0.01)$ - but the effect is small; reducing $\sigma$ by 0.1 reduces $\Delta M_{\bullet}$ by $\sim0.08$. This degeneracy was also found by \citet{lisync21}. 

Next, we replace an SMBH mass offset with cosmologically coupled BH growth \citep{farja07,cro19,croker2022well,cadoni23b}. In such models, the mass of the BH increases with the cosmological scale factor, due to being embedded in an expanding universe \citep{faraoni24}. The initial and final BH mass are related by:

\begin{equation}\label{eq:cosmocoup}
M_{\bullet,f} = M_{\bullet,i}\left(\frac{a_{f}}{a_{i}}\right)^{k}
\end{equation}

\begin{figure*}
\begin{center}
\includegraphics[width=0.98\linewidth]{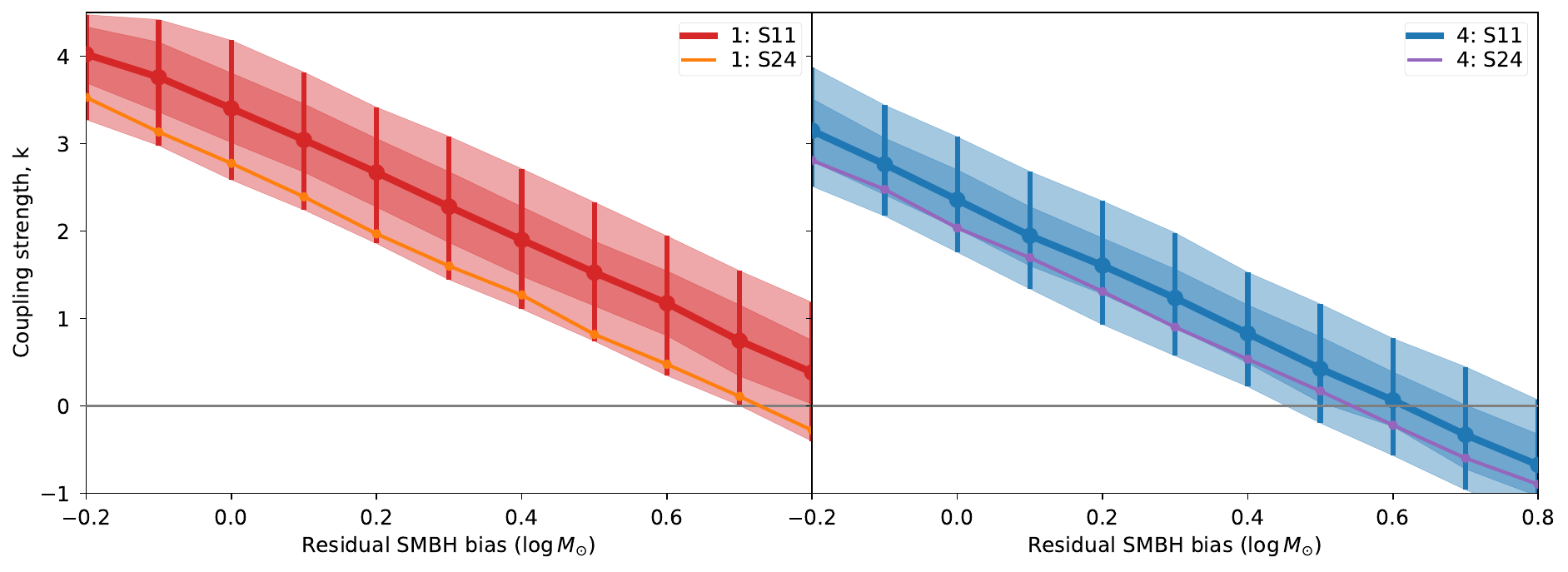}  
\caption{As Figure \ref{fig:offsetresidual}, but replacing an SMBH mass offset with the strength, $k$, of cosmological coupling (Equation \ref{eq:cosmocoup}) needed to align the high- and low-redshift relations.}
\label{fig:offsetoupling}
\end{center}
\end{figure*}

\noindent \citep{cro20} where $k$ is the coupling strength and $a$ is the cosmological scale factor. Since we are now fitting for $k$, we modify our analysis as follows. For a given $k$, we individually shift the low-redshift objects to $z=0$ and derive the resulting $M_{\bullet}-M_{*}$ relation. We then shift the intrinsic high-redshift relation to $z=0$ and compare it to the low-redshift relation. The results are shown in Figure \ref{fig:offsetoupling}.  With set \#1, the S11 (S24) SEV calibrator, and no residual SMBH bias, $k=3.40^{+0.78}_{-0.82}$ ($k=2.72^{+0.87}_{-0.56}$). With set \#4, $k=2.32^{+0.75}_{-0.57}$ ($k=1.94^{+1.27}_{-0.99}$). The scatter is comparable to that found for the $\Delta M_{\bullet}$ fits, and we see the same, small, degenracy: $(\tau,p)=(0.14,<0.01)$ (decreasing $\sigma$ by 0.1 will change $k$ by at most 0.5).

\section{Discussion}\label{sec:discuss}

\subsection{Selection bias in high-redshift AGN}
The selection bias in the high-redshift sample depends on the SMF, BHMF, and ERDF. Different choices can give comparable observed relations, but intrinsic relations that differ by up to $\sim0.4$ in slope and $\sim0.5$\,dex in intercept (Figure \ref{fig:vsbtest1}). The variation in the SMF, BHMF, and ERDF that give this range in the intrinsic relation are comparable to those  seen between $\Delta z\sim0.5$ intervals over $0<z \lesssim 4$ (S15, M21, W23).  Our results also show that constraints on the SMF, BHMF, and ERDF from samples of order $10^{2}$ objects in the $M_{\bullet}-M_{*}$ plane are challenging. Furthermore, the choice of SEV calibration can shift the intrinsic relation by $\sim0.2$\,dex, and alter its slope (Figure \ref{fig:vsbtest1}). 

There are two further sources of uncertainty. First is the virial factor, $f_{vir}$, SMBH masses from the virial method are computed via: $M_{\bullet} = f_{vir}(c \tau v^{2}/G)$, in which $c\tau$ is the time delay, $v$ is the line velocity width, and $G$ is Newton's constant. We have assumed $f_{vir}=4.3$, but this value is debated \citep{mejia18}. Some studies find $2\lesssim f_{vir}\lesssim 4$ \citep{shank19,mandal21,shen23rmkey}, which would lower our high-redshift SMBH masses by $\sim0.3$\,dex. Other studies argue for $f_{vir}\sim1-20$, and that it can depend on accretion rate \citep{liu22,liu24}. If true then $\langle f_{vir}\rangle\sim4$ may be reasonable, but a dependence on accretion rate may affect the slope and intercept of the intrinsic $M_{\bullet}-M_{*}$ relation, akin to the effect of different SEV calibrators. Second is $p_{act}$. We have assumed a loose range of $0.0001 < p_{act} < 0.2$ over all SMBH masses.  It is however possible that a tighter range, or one that depends on SMBH mass, could alter the values of $a$, $b$, and $\sigma_{t}$ that give acceptable observed relations. 

A thorough investigation of these dependencies would require a much larger sample. We argue though that any study of the $M_{\bullet}-M_{*}$ relation in AGN requires a well-matched measure of the SMF, BHMF, and ERDF to minimize bias.  Even then, residual uncertainty from the choice of SEV calibrator, $f_{vir}$, and $p_{act}$, will lead to a systematic uncertainty  in the $M_{\bullet}-M_{*}$ relation of $\sim0.3-0.5$\,dex.  The $M_{\bullet}-M_{*}$ plane is thus a challenging arena to study galaxy assembly with redshift. The $M_{\bullet}-$ stellar velocity dispersion plane may be more fundamental, and less prone to bias \citep{shank17}.

\subsection{ETG assembly over $0<z<1$}
\begin{figure}
\begin{center}
\includegraphics[width=0.98\linewidth]{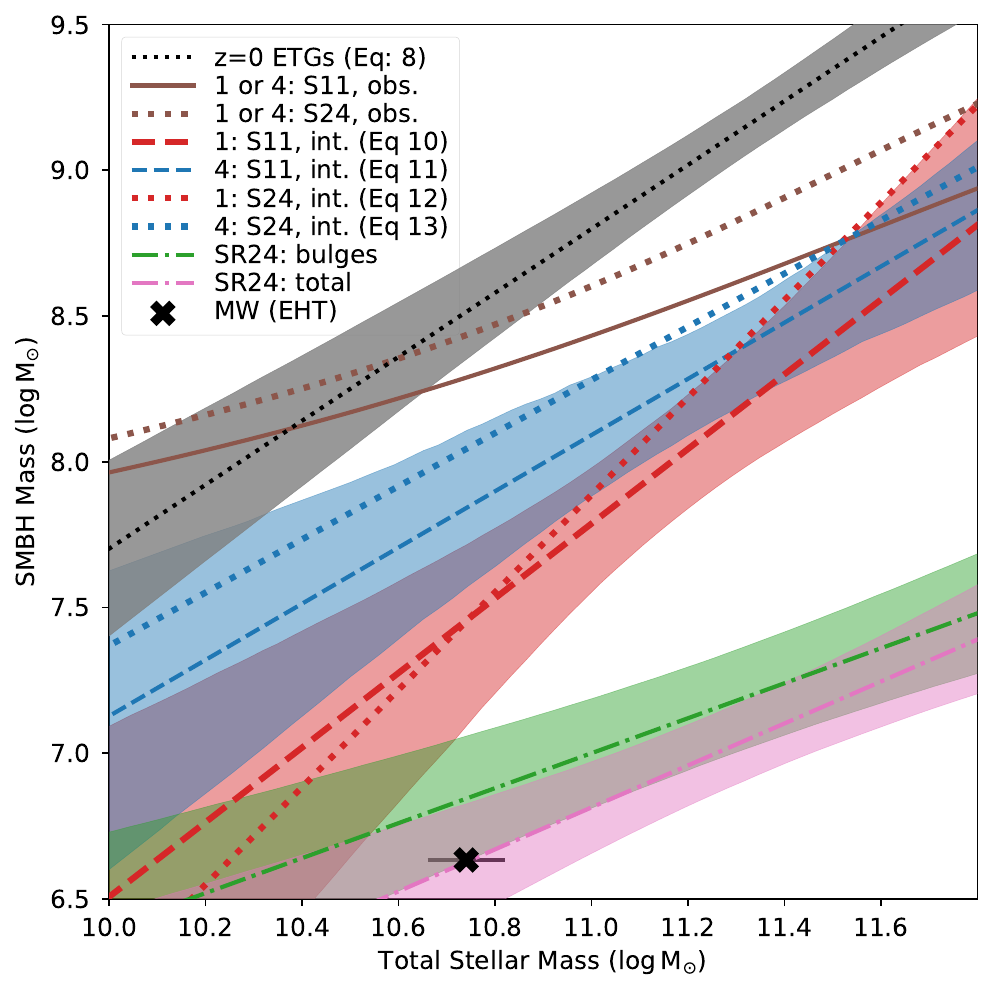} 
\caption{Our low- and high-redshift relations compared to the $z\sim0$ AGN relations from \citet[][who use the \citet{reines13} SEV relation]{sturmreies24}. Also shown is the Milky Way with SMBH mass from the \citet{ehtmwmass22}. We do not plot other literature relations  \citep[e.g.][]{greene16,bentz18,schutte19,benn21,liscale23}  as their uncertainties are significant, but in general they are consistent with ours.}
\label{fig:lhplanes2}
\end{center}
\end{figure}

If our high-redshift sample are ETGs emerging onto the red sequence, then our high- and low-redshift $M_{\bullet}-M_{*}$ relations can diagnose the processes that affect ETG assembly between $z\sim1$ and $z\sim0$. From Figure \ref{fig:lhplanes2} our high-redshift relation lies below our low-redshift relation but above the  relation for $z\sim0$ AGN (albeit for lower luminosity AGN in all host types). We here examine the implications of these offsets, and the possible further bias corrections (\S\ref{sec:discbias}) and physical mass changes (\S\ref{sec:discphys}) that may align them.

\subsubsection{Alignment with bias corrections}\label{sec:discbias}
Depending on parameter choices, $\sim0.4-1$\,dex of bias corrections, beyond those we applied, are required to align the high-redshift intrinsic relation with the low-redshift relation.  A plausible origin for some of this is dynamical selection bias in the low-redshift sample.  From \citet{shank16a}, dynamical selection bias much above $0.4$\,dex is unlikely in the stellar mass range of our sample, but a bias of up to $0.4$\,dex is feasible. There is support for this idea from studies of the redshift evolution of scaling relations by integrating along curves of $M_{*}$ \citep{yang18,shankar20,carraro20}, which find that, with dynamical selection bias in low-redshift objects, the local $M_{\bullet}-M_{*}$ relation can be reconciled with high-redshift measurements. Dynamical selection bias in low-redshift ETGs has, however, not been directly confirmed.

\begin{figure}
\begin{center}
\includegraphics[width=0.98\linewidth]{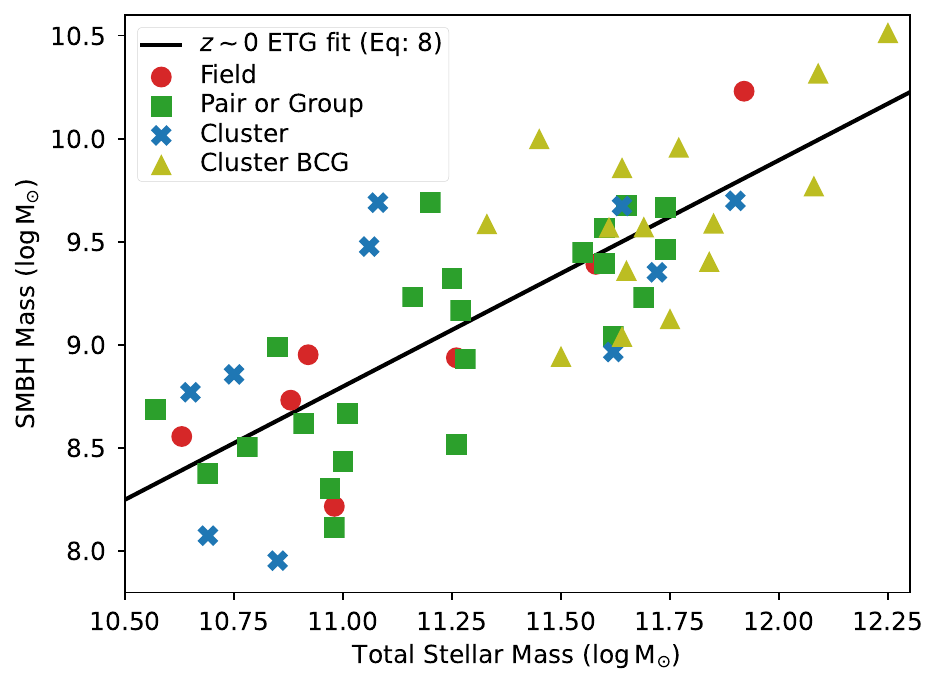} 
\caption{As Figure \ref{fig:lhplanes} left, with the sample classified by their local environment. The sample are all at $z<2$, so residual redshift bias will be insignificant.} 
\label{fig:localenvs}
\end{center}
\end{figure}

Another selection bias that may align the high- and low-redshift relations is ``downsizing," in which ETGs with relatively more massive SMBHs complete their assembly earlier. In this scenario, our $z\sim0.8$ ETGs represent those that complete their assembly relatively late, and hence have undermassive SMBHs. Support for this idea comes from studies of high-redshift starbursts \citep{wafa13,wilk17} and of overmassive SMBHs at $z\gtrsim2$ \citep[e.g.][]{maiolino24,furt24}, though these studies do not restrict to ETG hosts. If interpreted this way, our results strongly favor downsizing as they place the $z\sim0.8$ AGN in ETG hosts between our low-redshift and $z>1$ $M_{\bullet}-M_{*}$ relations. A decisive test of this possibility requires knowledge of the stellar mass assembly histories of our sample, which only a few have \citep{mcder15}. We can however perform a basic check by first noting that ETGs assemble earlier in denser environments \citep{thomas05}. If low-redshift ETGs with higher $M_{\bullet}/M_{*}$ ratios assembled earlier, then they may reside in denser environments. Our low-redshift sample lack homogeneous measures of environment, so we classify them as isolated, group, cluster, or cluster BCG, based on the information in the NASA/IPAC Extragalactic Database, and plot them in the $M_{\bullet}-M_{*}$ plane (Figure \ref{fig:localenvs}).  There is no clear trend for systems with {\itshape relatively} more massive SMBHs to prefer denser environments. This test is not robust, but it does suggest that any signatures of downsizing in ETG assembly are subtle. 

A further possible bias is if the high-redshift hosts are contaminated by dusty lenticular galaxies. The possibility of such a population has been raised by \citet{graham24}. Such a population, with on average smaller SMBHs, could explain our result.  Based on the selection of the AGN to have low reddening, we do not believe this is likely, but we cannot exclude it.  

 Finally, there are two measurement biases that could help align our relations. First is residual stellar mass measurement bias. With the S24 calibrator, the required stellar mass bias would be $\sim0.4$\,dex. This seems only barely plausible, for the reasons given in  \S\ref{sec:prelim2}. It cannot be discounted though. The high-redshift host SEDs must be resolved against a much brighter AGN, meaning that a significant overestimate in $M_{*}$ is possible. We lack the data to test this. However, if the stellar masses of the high-redshift sample are corrected downward, then this will increase the SMBH selection bias in the sample.  For example, a $0.3$\,dex downward $M_{*}$ correction would increase the SMBH selection bias by $\sim0.2$\,dex.  Stellar mass measurement bias thus seems an implausible route to aligning the high- and low-redshift relations. Second, our high-redshift host classifications may be incorrect. If they are contaminated by disk hosts, then the offset we observe between the high- and low-redshift relations could signpost an episode of preferential SMBH growth in a major merger \citep[e.g.][]{farrah16,farrah22}. Testing this possibility would require spatially resolved imaging of the quasar hosts, which we do not have.

\subsubsection{Alignment with mass changes}\label{sec:discphys}
Either stellar mass loss or preferential SMBH mass growth  between $z\sim0.8$ and $z=0$ ($\Delta t\sim7$ Gyr) could align the high- and low-redshift relations. Stellar mass loss is implausible, since processes such as cluster infall cannot remove enough stellar mass to align the relations (F23, and references therein). This leaves preferential SMBH mass growth. Of order $5\times10^{8}$M$_{\odot}$ of SMBH growth is required to align the relations, assuming no increase in stellar mass. Four routes could achieve this: ``slow and steady" accretion, rapid accretion as an AGN, SMBH mergers, or cosmological coupling.

For slow and steady accretion to align the relations, the average accretion rate is $\dot{M}_{\bullet}\sim7\times10^{-2}$M$_{\odot}$yr$^{-1}$.  While  there are routes that could supply gas at such rates \citep[e.g.][]{fabian24,ivey24}, and the typical $\dot{M}_{\bullet}/\dot{M}_{*}$ ratio in ETGs \citep{mccdonald21} gives a plausible SFR, the accretion rate itself is typical of AGN. This seems unlikely. Brief, rapid accretion is more viable, since high accretion rates can be triggered in gas-rich mergers, but would require most ETGs to go through at least one such merger between $z\sim0.8$ and $z=0$. The merger rate required seems in tension with recent predictions \citep[e.g.][]{oleary21}.

For SMBH-SMBH mergers to align the relations then, if exotic possibilities \citep{presstek72} are neglected, the $M_{\bullet}/M_{*}$ ratio in the `other' galaxy must be at least that seen in local ETGs.  This is in principle plausible, as ultracompact galaxies with overmassive SMBHs have been found in the haloes of ETGs \citep{seth14,voggel18}. However, to increase the SMBH masses by at least 0.3 dex relative to the stellar mass would require tens of such mergers. 

Finally, cosmologically coupled BH growth could align the high- and low-redshift relations. This requires $k \simeq 1-3$, depending on the choices of SMF, BHMF \& ERDF parameters, and SEV calibrator. Values in this range are plausible; $k=3$ has been proposed to link BHs with dark energy \citep{farrah23de}, and $k=1$ has a known exact solution \citep{cadoni23}. Neither value is, however, independently preferred. \citet{lacy24} have shown that, with conventional assumptions about the local SMBH mass density \citep[e.g.][]{grahdriv07}, $k<2$ is required, but with the SMBH mass density proposed by \citet{nanograv23} then $k>2$ may be required. Some studies of stellar-mass BHs are consistent with $k\sim3$ \citep{gaoli23}  but others find that $k\gtrsim2$ may require the zero-age masses of BHs formed through stellar collapse to lie below the  Tolman-Oppenheimer-Volkov limit\footnote{This is possible, since the TOV limit is only a lower bound on BH mass if BHs have trapping surfaces, which cosmologically coupled BHs may not have. To date though, no BH below the TOV limit has been found.} \citep{rodriguez23,andrae23,ghodla23,amendola24,mlinar24}. Our study cannot resolve this controversy, but does show that $k>0$ is viable from observations of galaxy scaling relations.

\section{Conclusions}\label{sec:conc}
We have studied the $M_{\bullet}-M_{*}$ relation in two samples of ETGs with $M_{*}>10^{10.5}$M$_{\odot}$. First, quiescent, morphologically-selected ETGs at $z<0.2$ with SMBH masses from dynamical modelling. Second, AGN (L$_{bol} \geq 10^{45.5}$ erg s$^{-1}$) in ETGs at $z\sim0.8$, selected through SED fitting, with SMBH masses from the virial method. We compare their $M_{\bullet} - M_{*}$ relations to constrain the assembly histories of ETGs. We find:

1 - For the $z<0.2$ sample the observed mean relation is $\log{M_{\bullet}} = 8.80^{+0.12}_{-0.12} + 1.10^{+0.21}_{-0.21} (\log{M_{*} - 11})$, with scatter $\sigma = 0.37^{+0.08}_{-0.06}$. This is in line with prior work, and suggests a tight intrinsic relation. For the $z\sim0.8$ AGN, we adopt the \citet{shen11cat} virial relation, $f_{vir}=4.3$, and correct for selection bias. The intrinsic mean relation is then $\log{M_{\bullet}} = 7.80^{+0.18}_{-0.24} + 1.25^{+0.68}_{-0.47} (\log{M_{*} - 11})$ with $\sigma = 0.24^{+0.18}_{-0.38}$.  This relation lies below our $z<0.2$ relation in SMBH mass by $\sim1$\,dex, but has a consistent slope. It lies above the $M_{\bullet}-M_{*}$ relation for local AGN by $\sim0.5$\,dex. 

2 - The $z\sim0.8$ intrinsic $M_{\bullet}-M_{*}$ relation depends on the SMF and active BHMF \& ERDF used to calculate the selection bias, the virial relation, the virial factor, and the active fraction. Different choices of these variables can alter the  intercept of the intrinsic relation by up to $\sim0.6$\,dex, and the slope by up to $\sim0.4$. These changes can arise by altering the SMF, BHMF, and ERDF by amounts comparable to changes in them seen in redshift intervals of $\Delta z\sim0.5$.  We further show that the SMF and BHMF/ERDF cannot be constrained using samples of order $10^{2}$ objects in the $M_{\bullet}-M_{*}$ plane. Since the virial relation, virial factor, and active fraction are also challenging to constrain with such samples, these variables act as systematic biases in any study of the $M_{\bullet}-M_{*}$ relation in AGN. 

3 - Reasonable choices of the model variables gives an SMBH mass offset between the $z<0.2$ and $z\sim0.8$ relations of $\sim0.4-1$\,dex. We explore ways to align the two $M_{\bullet} - M_{*}$ relations. Dynamical selection bias in the $z<0.2$ sample could account for some, possibly all, of the offset. Another possibility is ``downsizing"; ETGs with relatively more massive SMBHs assemble at higher redshifts.  Other possibilities that could align the relations include brief, rapid accretion events, contamination by disk or dusty lenticular hosts, or cosmologically coupled SMBH mass growth. 

4 - Our results demonstrate the importance of several avenues of study for using the $M_{\bullet}-M_{*}$ plane as a diagnostic tool. These include the redshift, environment and morphology dependence of the SMF, BHMF, and ERDF, the calibration of stellar and virial SMBH masses, and the possible model dependence of the virial factor and active fraction.

\begin{acknowledgments}
We thank the referee for a very helpful report. We thank Michael Rowan-Robinson, River, and John Petty for helpful discussions.  FS acknowledges partial support from the EU H2020-MSCA-ITN-2019 Project 860744 ‘BiD4BESt: Big Data applications for black hole Evolution STudies’ (Coordinator: F. Shankar). RAW acknowledges support from NASA JWST Interdisciplinary Scientist grants NAG5-12460, NNX14AN10G and 80NSSC18K0200 from GSFC. JA acknowledges financial support from the Science and Technology Foundation (FCT, Portugal) through research grants PTDC/FIS-AST/29245/2017, UIDB/04434/2020 (DOI: 10.54499/UIDB/04434/2020) and UIDP/04434/2020 (DOI: 10.54499/UIDP/04434/2020).
\end{acknowledgments}

\bibliography{ellipcc}{}
\bibliographystyle{aasjournal}

\end{document}